\documentclass[aps,prx,preprint,letterpaper,groupedaddress,longbibliography]{revtex4-2}
\usepackage{amsmath,amsthm,amsfonts}
\theoremstyle{definition}

\usepackage{multirow}
\usepackage{array}
\usepackage{soul}
\usepackage{float}
\usepackage{graphicx}
\usepackage{afterpage}
\usepackage{tabularx}

\usepackage[font=small,labelsep=quad,format=plain]{caption}
\usepackage{subcaption}
\usepackage[colorlinks=true,citecolor=blue,linkcolor=blue,urlcolor=blue]{hyperref}
\usepackage{cleveref}
\crefname{equation}{Eq.}{Eqs.}
\crefname{section}{Sec.}{Secs.}
\crefname{subsection}{Sec.}{Secs.}
\crefname{appendix}{Appendix}{Appendices}
\crefname{figure}{Fig.}{Figs.}
\crefname{table}{Table}{Tables}
\crefname{proposition}{}{}
\crefname{corollary}{}{}

\usepackage{bm}
\usepackage{physics}
\usepackage[dvipsnames]{xcolor} 

\newcommand{\eunits}{kV cm\textsuperscript{-1}}
\newcommand{\smalleunits}{V cm\textsuperscript{-1}}

\newcommand{\qnumber}{\lambda}
\newcommand{\qnumberp}{\lambda'}
\newcommand{\qnumbertext}{\lambda}

\begin{document}

\title{Hot hole transport and noise phenomena in silicon at cryogenic temperatures from first principles}

\author{David S. Catherall}
\affiliation{Division of Engineering and Applied Science, California Institute of Technology, Pasadena, CA 91125, USA}
\author{Austin J. Minnich}
\email{aminnich@caltech.edu}
\affiliation{Division of Engineering and Applied Science, California Institute of Technology, Pasadena, CA 91125, USA}
\date{\today}

\begin{abstract}
The transport properties of hot holes in silicon at cryogenic temperatures exhibit several anomalous features, including the emergence of two distinct saturated drift velocity regimes and a non-monotonic trend of the current noise versus electric field at microwave frequencies. Despite prior investigations, these features lack generally accepted explanations. Here, we examine the microscopic origin of these phenomena by extending a recently developed ab-initio theory of high-field transport and noise in semiconductors. We find that the drift velocity anomaly may be attributed to scattering dominated by acoustic phonon emission, leading to an additional regime of drift velocity saturation at temperatures $\sim 40$ K for which the acoustic phonon occupation is negligible; while  the non-monotonic trend in the current noise arises due to the decrease in momentum relaxation time with electric field. The former conclusion is consistent with the findings of prior work, but the latter distinctly differs from previous explanations. This work highlights the use of high-field transport and noise phenomena as sensitive probes of microscopic charge transport phenomena in semiconductors.

\end{abstract}

\maketitle
\newpage

\section{Introduction}

The microscopic processes underlying charge transport in semiconductors are of fundamental and practical interest \cite{Lundstrom:2000,Ferry:2000,Ponce:2020,Bernardi:2016,Giustino:2017}. Numerical approaches to study transport phenomena have typically employed Monte Carlo simulations \cite{Lundstrom:2000}, which are capable of treating realistic device geometries \cite{jacoboni:1989,Mateos:2000} and have been extended to full-band simulators \cite{Hess:1991,Jungemann:1999}. Within the last decade, the development of the ab-initio treatment of the electron-phonon interaction has enabled the calculation of transport properties in homogeneous systems without any fitting parameters \cite{Bernardi:2016,Giustino:2017}. Recent findings with these methods include the importance of 2ph scattering in GaAs \cite{Lee:2020, Cheng:2022} and Si \cite{Ben:2022} as well as the discovery of simultaneously high electron and hole mobilities in BAs \cite{Liu:2018,Shin:2022,Yue:2022}. Most ab-initio calculations have been restricted to the low-field regime in which electrons are in thermal equilibrium with the lattice. However, outside of the low-field regime, phenomena such as field-dependent mobility, drift velocity saturation, and negative differential resistance may occur \cite{Jacoboni:1979,Ottaviani:1975,Canali:1975}. Furthermore, at high fields qualitatively new information can be obtained from noise quantities, such as the current fluctuation power spectral density (PSD), due to the breakdown of equilibrium relationships including the Einstein \cite{Reggiani:1985_2} and Nyquist equations \cite{Nougier:1980}. Recent studies have extended the ab-initio method to the high-field regime and applied them to both transport and noise in GaAs \cite{Choi:2021,Cheng:2022} and Si \cite{Catherall:2022,Ben:2022}.

Charge transport in \textit{p}-Si at cryogenic temperatures and microwave frequencies is of particular interest due to two anomalous transport characteristics. First, at large electric fields of tens of \eunits, semiconductors generally exhibit drift velocity saturation in which the drift velocity no longer increases with field \cite{Lundstrom:2000}. In \textit{p}-Si below 40 K, an additional regime of drift velocity saturation occurs at considerably smaller fields, approximately $0.1$ \eunits~\cite{Ottaviani:1975}. This feature has been studied numerically by semi-analytical \cite{VonB:1976,Ohno:1988} and full-band Monte-Carlo methods \cite{Fischer:2000}. While an initial investigation attributed the feature to the non-parabolic hole dispersion and Bose-Einstein distribution occupation of phonons \cite{VonB:1976}, a later study instead attributed the behavior solely to population of the spin-orbit band at high fields \cite{Ohno:1988}. The ``shoulder" feature has more recently been computationally reproduced using full-band Monte Carlo \cite{Fischer:2000}, but without analysis. Therefore, the explanation of the anomalous saturation regime remains unresolved.\par

A second anomalous feature appears in the PSD. It has been observed experimentally that in \textit{p}-Si at 10 GHz and 77 K, a non-monotonic (peak) trend is observed for current noise both longitudinal and transverse to an applied DC electric field. This feature has also been observed in \textit{n}-Si and Ge \cite{Bareikis:1980,Bareikis:1981}. The trend was originally ascribed to an initial increase due to carrier heating followed by a decrease due to a decreasing average momentum relaxation time \cite{Bareikis:1981}. In Sec.~9.3 of Ref.~\cite{Hartnagel:2001}, the trend was attributed to a field-dependent energy relaxation time which influences the convective noise mechanism (described in section 7.2 of the same reference). The convective mechanism arises from the coupling of velocity and energy fluctuations of the charge carriers, leading to a decrease in the PSD with increasing electric field for materials with 
a sublinear current-voltage characteristic. Given these differing explanations, the origin of the peak remains unclear.

Here, we investigate these phenomena by computing the drift velocity and PSD versus field of hot holes in silicon between 6 and 77 K using a modified implementation of the high-field ab-initio method which is applicable at cryogenic temperatures. We show that the additional drift velocity saturation regime arises from scattering dominated by acoustic phonon emission. For the PSD, we find that the peak occurs due to the decrease of the average momentum relaxation time with increasing electric field, an explanation not given in prior literature. This work demonstrates the use of high-field first-principles transport and noise calculations as useful tools in investigating the microscopic phenomena in charge transport.

\section{Theory and numerical methods}

\subsection{Overview}

The methods used in this work have been described previously \cite{Choi:2021, Cheng:2022, Catherall:2022}. Briefly, the transport of charge carriers due to an applied electric field may be described by the Boltzmann transport equation (BTE). For a spatially homogeneous system, the BTE is given by

\begin{equation}
    {
        \frac{q\mathbf{E}}{\hbar}
        \cdot
        \nabla_\mathbf{k}
        f_{\qnumber}
        =
        -\Theta_{\qnumber\qnumberp}
        \Delta f_{\qnumberp}
    }
    \label{shortBTE}
\end{equation} 

\noindent where $q$ is the carrier charge, $\mathbf{E}$ is the electric field vector, $f_{\qnumber}$ is the carrier occupation function indexed by $\qnumbertext$ which represents the combined indices of band $n$ and wave vector $\mathbf{k}$, and $\Delta f_{\qnumber}$ represents the deviation from equilibrium as $\Delta f_{\qnumber}=f_{\qnumber}-f^0_{\qnumber}$ where $f^0_{\qnumber}$ is the Fermi-Dirac distribution at a specified lattice temperature and chemical potential.  $\Theta_{\qnumber\qnumberp}$ is the linearized collision integral as given by Eq.~3 of Ref.~\cite{Choi:2021}. The collision integral depends on the phonon populations which may be perturbed by carrier scattering, but owing to the small free carrier densities in the relevant experiments ($\lesssim10^{14}\,\mathrm{cm}^{-3}$), this effect may be neglected here \cite{Ottaviani:1975}.


Equation \ref{shortBTE} can be expressed as a linear system of equations using the linearized collision integral and a finite difference approximation for the reciprocal space gradient, as described in Sec.~II.A. of Ref.~\cite{Choi:2021}. The only modification to the governing equations in the present work is a change of sign for the reciprocal-space gradient to account for the charge carriers being holes. The BTE then takes the form

\begin{equation}
    {
        \sum_{\qnumberp}
        \Lambda_{\qnumber\qnumberp}
        \Delta f_{\qnumberp}
        =\sum_\gamma
        \frac{e\mathrm{E}_\gamma}{k_BT}
        v_{\qnumber,\gamma}
        f^0_\qnumber
    }
    \label{fullBTE}
\end{equation}

\noindent where $\mathrm{E}_\gamma$ and $v_{\qnumber,\gamma}$ are the electric field strength and hole velocity in the $\gamma$ Cartesian axis, $k_B$ is the Boltzmann constant, and $T$ is the lattice temperature. The relaxation operator $\Lambda_{\qnumber\qnumberp}$ is defined as

\begin{equation}
    {
        \Lambda_{\qnumber\qnumberp}
        =
        \Theta_{\qnumber\qnumberp}
        -\sum_\gamma
        \frac{e\mathrm{E}_\gamma}{\hbar}
        D_{\qnumber\qnumberp,\gamma}
    }
    \label{relop}
\end{equation}

\noindent where $D_{\qnumber\qnumberp,\gamma}$ is the finite difference matrix (FDM) approximating the $\gamma$-axis component of the reciprocal-space gradient. The solution to the linear system, $\Delta f_{\qnumber}$, can be used to calculate  various observables including the mobility and PSD. The mobility is given as \cite{Li:2015}

\begin{equation}
    {
        \mu_{\alpha\beta}(\mathbf{E})
        =
        \frac{2e^2}{k_BT\mathcal{V}}
        \sum_{\qnumber}
        v_{\qnumber,\alpha}
        \sum_{\qnumberp}
        \Lambda^{-1}_{\qnumber\qnumberp}
        \left(
            v_{\qnumberp,\beta}
            f^0_{\qnumberp}
        \right)
    }
    \label{mobility}
\end{equation} 

\noindent where $\mathcal{V}$ is the supercell volume, $\alpha$ is the direction along which the current is measured, and $\beta$ is the direction along which the electric field is applied. The PSD is given as

\begin{equation}
    {
        S_{j_\alpha j_\beta}(\mathbf{E},\omega)
        =
        2\left(
            \frac{2e}{\mathcal{V}}
        \right)^2
        \mathfrak{R}
        \left[
            \sum_{\qnumber}
            v_{\qnumber,\alpha}
            \sum_{\qnumberp}
            \left(
                i\omega\mathbb{I}
                +\Lambda
            \right)_{\qnumber\qnumberp}^{-1}
            \left(
                f^s_{\qnumberp}
                \left(
                    v_{\qnumberp,\beta}
                    -V_\beta
                \right)
            \right)
        \right]
    }
    \label{vPSD}
\end{equation}
\noindent where $\omega$ is the angular frequency, $\mathbb{I}$ is the identity matrix, and  $f^s_{\qnumber}=f^0_{\qnumber}+\Delta f_{\qnumber}$ is the steady distribution. Here, $V_\beta$ is the drift velocity, given by
\begin{equation}
    {
        V_\beta
        =
        \frac{1}{N}
        \sum_{\qnumber}
        v_{\qnumber,\beta}
        f^s_{\qnumber}
    }
    \label{aveV}
\end{equation}

\noindent where $N=\sum_{\qnumber}f_{\qnumber}$ is the number of holes in the Brillouin zone. Further details are available in Sec.~II.B. of Ref.~\cite{Choi:2021}.

\subsection{Finite difference matrix}
Within the Monkhorst-Pack  Brillouin zone (BZ) discretization scheme used here \cite{Monkhorst:1976}, the reciprocal-space gradient may be approximated by an FDM as defined by Eqns. 8 and B2 of Refs.~\cite{Wannier90,Marzari:1997}, respectively. In prior high-field works, \cite{Choi:2021,Cheng:2022,Catherall:2022,Ben:2022} the inclusion of only first-nearest-neighbors provided adequate accuracy at $T\geq 77$ K. At temperatures below 77 K, we found the quality of this numerical finite difference approximation to be poor. This issue could, in principle, be addressed by increasing the grid density. However, for $T<77$ K the grid density cannot be made sufficiently large to achieve the needed accuracy due to computational limitations.\par

Instead, we utilize the same prior FDM method but use multiple shells for the finite difference approximation, as originally proposed in Ref.~\cite{Marzari:1997}. To determine the shell weights, an analytical method cannot be used since the combination of a BCC Monkhorst-Pack mesh for Si and the use of multiple shells leaves low-order mixed partial derivatives which cannot be eliminated using a finite number of shells. To overcome this limitation, we solve for shell weights for a selected electronic band by the minimization of an objective function measuring the difference between exact and approximate derivatives

\begin{equation}
    \mathrm{Error}(T,\{w_b\})
    =
    \frac
        {\sum_\lambda \left( \left[\frac{\partial f^0_\lambda}{\partial k_\gamma}\right] - [D_\gamma(\{w_b\})\,f^0_\lambda]\right)^2 f^0_{\lambda}}
        {\sum_\lambda \left[\frac{\partial f^0_{\lambda}}{\partial k_\gamma}\right]^2 f^0_\lambda}
    \label{obj_eqn}
\end{equation}

\noindent where $\left[\frac{\partial f^0_{\lambda}}{\partial k_\gamma}\right]$ is the analytical derivative of the Fermi-Dirac function $f^0_{\lambda}$ and $D_\gamma(\{w_b\})$ is the FDM representation of a single $\gamma$-axis component of the reciprocal-space gradient. The temperature-dependence of the objective function arises from the temperature-dependence of $f^0_{\lambda}$.  The FDM is defined using $\{w_b\}$, which is the set of shell weights and are the minimization parameters, subject to the constraint $\sum_b w_b = 1$. The subscript $b$ identifies the nearest-neighbor shell. We define the total number of shells in a scheme by $B$.\par

It was found that relative to the one-shell case $(B=1)$ at $T\leq 77$ K, the error defined by Eq.~\ref{obj_eqn} decreased by 96\% with the inclusion of three shells ($B=3$), and by a further 77\% with seven shells ($B=7$). In terms of transport properties, we compare the high-field DC mobility in different cases via RMS difference (defined as $||y_{1}-y_{2}||_2\,/\,||y_{1}||_2$, where $y$ is the property of interest). The mobility obtained using $B=3$ and $B=7$ agreed with the $B=1$ case to within 2.1\% even at 20 K and up to 10 \eunits. The RMS difference between $B=3$ and $B=7$ was 0.3\%. This agreement indicates that, despite being optimized to accurately calculate the gradient of the Fermi-Dirac function ${f^0_\lambda}$, the higher-order FDM still provides an accurate result when applied to the hot carrier distribution function $f_\lambda$.\par

\begin{table}
    \newcolumntype{Y}{>{\centering\arraybackslash}X}
    \begin{tabularx}{14cm}{>{\centering}p{3cm} YYYYYYY}
        \hline\hline
         & \multicolumn{7}{c}{Shell ($b$)} \\
        \cline{2-8}
        Total shells ($B$) & 1 & 2 & 3 & 4 & 5 & 6 & 7 \\
        \hline
        1 & 1.000 &       &       &       &       &       & \\
        2 & 0.725 & 0.275 &       &       &       &       & \\
        3 & 1.390 & 0.530 &-0.920 &       &       &       & \\
        4 & 1.270 & 0.600 &-0.490 &-0.380 &       &       & \\
        5 & 1.280 & 0.590 &-0.480 &-0.370 &-0.020 &       & \\
        6 & 1.310 & 0.590 &-0.420 &-0.680 &~0.030 & 0.170 & \\
        7 & 1.380 & 0.730 &-0.330 &-0.980 &-0.450 & 0.100 & 0.550 \\
        \hline\hline
    \end{tabularx}
    \caption{Calculated FDM shell weights $w_b$ using \cref{obj_eqn} and $T=77$ K for various numbers of shells $B$.}
    \label{tab:shell_weights}
\end{table}

For noise properties, however, $B\geq3$ was found to be necessary to remove numerical artifacts. Therefore, we used $B=3$ at 77 K and $B=7$ for $T < 77$ K. For points near the edge of the defined grid for which the default number of shells ($B$) is not possible, we used the maximum number of shells with fully defined neighboring points. For uniformity, the shell weights were calculated using $T=77$ K, as temperature had only a negligible effect on $\{w_b\}$. For example, calculating $\{w_b\}$ for $B=7$ at $T=20$ K results in an RMS difference in the 20 K drift velocities of only 3.7\% compared to calculations using shell weights optimized at 77 K. Shell weights $w_b$ for each value of $b$ and $B$ are as shown in \cref{tab:shell_weights}.\par

\subsection{Numerical details}

Band structure and phonon calculations were performed in Q\textsc{uantum} ESPRESSO \cite{Giannozzi:2009} as in Ref.~\cite{Catherall:2022}. Fine-grid interpolation and scattering rate calculations were performed via P\textsc{erturbo} \cite{Zhou:2021}. A $250^3$ ($250 \times 250 \times 250$) grid with 2.5 meV Gaussian smearing were used for $T>6$ K, and $325^3$ grid with 1.25 meV smearing for $T=6$ K. Increasing the Gaussian width to 1.55 meV at 6 K resulted in a maximum change in drift velocity of 1.7\% at any field. The energy windows for scattering rate calculations were 80 and 57 meV for $T>6$ K and $T = 6$ K, respectively. Increasing the energy window to 90 meV at 77 K and a grid of $180^3$ resulted in a 3\% difference at the maximum field studied (10 \eunits), and increasing the window to 70 meV at 6 K and a grid of $200^3$ resulted in a similar 3.5\% difference at 0.84 \eunits. The linearized Boltzmann transport equation \cite{Choi:2021} is then solved via the GMRES algorithm \cite{Scipy} to calculate the steady-state distribution function, mobility, and noise quantities \cite{Choi:2021,Catherall:2022}. Only the heavy holes were included to reduce computational costs; it was found that at 77 K omitting the light hole band resulted in an RMS difference of only 1.5\% compared to results with both the heavy and light holes, and this error decreases with decreasing temperature owing to the decreasing occupation of the light and split-off bands.

\section{Results}

\subsection{Drift velocity}

\begin{figure}
    \centering{
        \includegraphics[width=3.4in, height=3.0in]{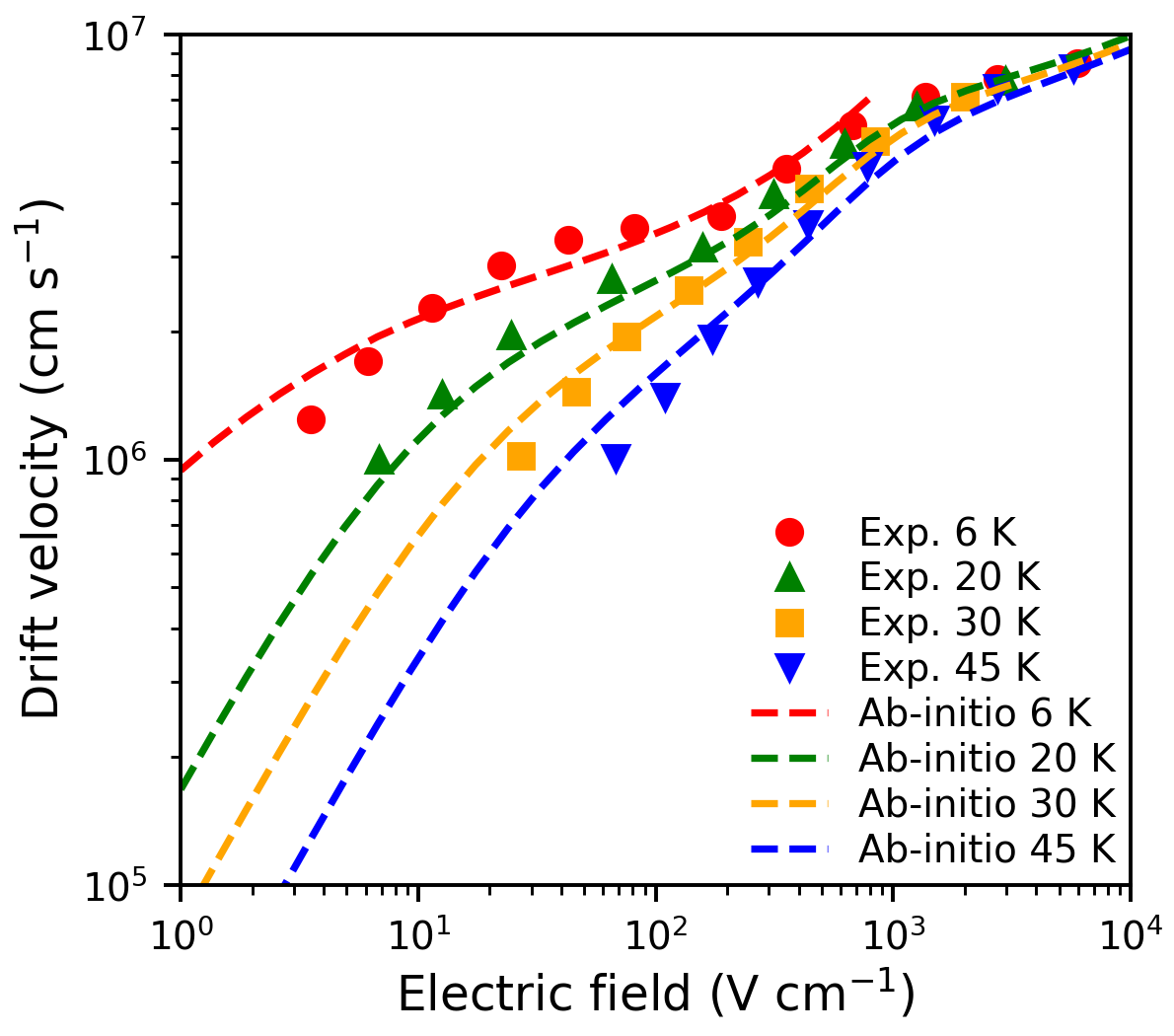}}
    \caption{Hole drift velocity versus electric field in the [100] direction. Experimental values from Ref.~\cite{Ottaviani:1975} (6 K as red circles, 20 K as green upward triangles, 30 K as yellow squares, 45 K as blue downward triangles). Calculated values also shown (dashed lines with identical color scheme to experiment).}
    \label{fig:drift_velocity}
\end{figure}

We first investigate the drift velocity characteristics of \textit{p}-Si. The calculated drift velocity-field curves at various cryogenic temperatures are shown in \cref{fig:drift_velocity}. We observe that the ab-initio calculations predict the two regimes of velocity saturation that are observed experimentally. At $\lesssim 10$ \smalleunits, the drift velocity increases linearly with field following the Ohmic trend, followed by an increase with lesser slope from 10 - 200 \smalleunits~corresponding to the anomalous saturation regime. At $\sim 200$ \smalleunits, the drift velocity increases further with nearly the same slope as in the Ohmic regime, but eventually enters another saturation regime at $\sim 500$ \smalleunits. Near-quantitative agreement with experiment is achieved to within 8\% (RMS) over all temperatures.\par

\begin{figure}
    \centering{
        \phantomsubcaption\label{fig:energy_scatter_a}
        \phantomsubcaption\label{fig:energy_scatter_b}
        \includegraphics[width=3.4in, height=5in]{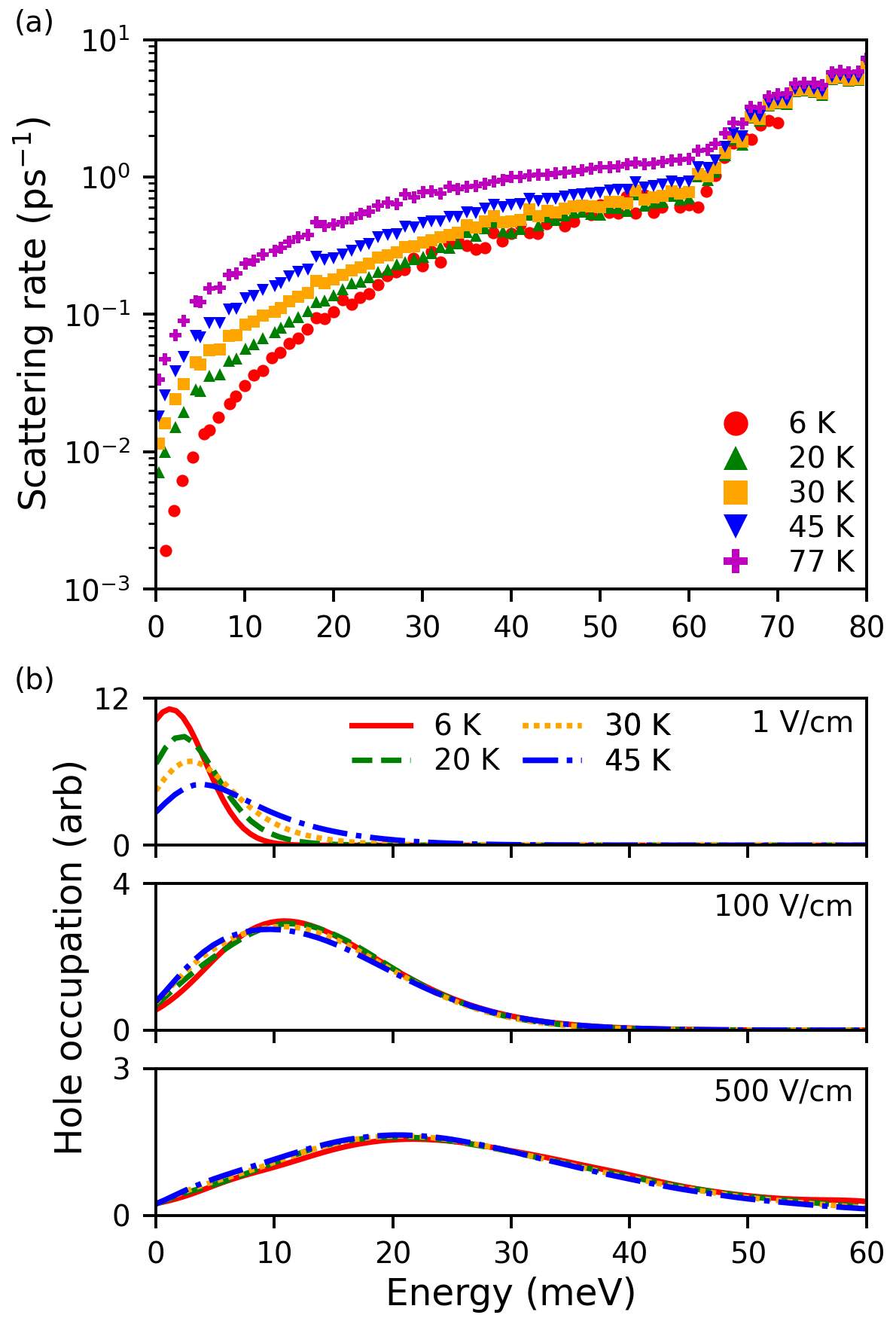}}
    \caption{(a) Calculated hole-phonon scattering rate versus energy from the valence band maximum at temperatures of 6 K (red circles), 20 K (green upward triangles), 30 K (yellow squares), 45 K (blue downward triangles), and 77 K (magenta crosses). (b) Hole occupation function versus energy from the valence band maximum at various electric fields applied in the [100] direction. The occupation is obtained using a kernel density estimate. Colors correspond to the same lattice temperatures as in (a).}
    \label{fig:energy_scatter}
\end{figure}

To gain insight into the origin of the low-field saturation regime, we show the calculated hole-phonon scattering rates versus energy in \cref{fig:energy_scatter_a}. In the following discussion, we focus only on the 20 K case for simplicity. At this temperature, we observe that the scattering rates increase rapidly with energy; from hole energies of 0 to 20 meV there is an increase in scattering rates by a factor exceeding 200. However, from 20 to 60 meV the rates are relatively constant, only rising by a factor of 4. Above 60 meV optical phonon emission becomes possible and the scattering rates again rapidly rise with energy.

These differing trends below and above $\sim 20$ meV arise due to two factors. First, below $\sim 40$ K, the population of zone-edge acoustic phonons rapidly diminishes with temperature. Consequently, the acoustic phonon absorption rate approaches zero while the acoustic phonon emission rate approaches a constant. Second, the phonon density of states, which increases quadratically up to a maximum at $\sim 20$ meV \cite{Kim:2015}, causes a corresponding increase in scattering rates with energy from $0-20$ meV. These factors together result in an increase in acoustic phonon emission from 0 to $\sim 20$ meV and relatively constant rates at higher energies up to $\sim 60$ meV.\par

We next examine the occupation function at various fields to establish which hole energies and corresponding scattering rates are relevant when the saturation feature occurs. We plot the hole occupation function versus energy and applied electric field in \cref{fig:energy_scatter_b}. At 1 \smalleunits, we observe that nearly all holes are confined to energies less than 10 meV, and the hole occupation exhibits a clear dependence on temperature. For sufficiently strong electric fields ($\gtrsim 10$ \smalleunits), the weight of the distribution shifts towards 10-30 meV as shown in \cref{fig:energy_scatter_b} at 100 \smalleunits, where the slope of drift velocity with field reaches a local minimum. In this energy range, we observe the rapid increase in scattering rates due to zone-edge acoustic phonon emission. At 500 \smalleunits, shown also in \cref{fig:energy_scatter_b}, the hole distribution has weight in the 30-60 meV region, where scattering rates are relatively constant. Above this field, some holes have energy exceeding  $\sim 60$ meV and may undergo optical phonon emission processes.\par

We may now explain the relationship between acoustic phonon scattering and the first regime of drift velocity saturation. At low fields ($\lesssim 10$ \smalleunits), the drift velocity is linear with field corresponding to the Ohmic regime, as holes are approximately in equilibrium with the lattice and the mean scattering rate remains independent of field, a trend which occurs at all temperatures. However, at cryogenic temperatures and high fields ($\sim 40-200$ \smalleunits), carrier heating shifts the occupation to energies where there is a large increase in scattering rates. Then, the increase in scattering rates with energy is far larger than the corresponding increase in band velocity, causing a decrease in mobility and the drift velocity saturation.\par

At higher fields, above $\sim 200$ \smalleunits, the increasing band velocity with energy compensates the relatively smaller increase in scattering rates. Therefore, the saturation ``shoulder" ends, and drift velocity again increases with increasing field. Finally, at about $\sim 500$ \smalleunits, optical phonon emission begins and another regime of drift velocity saturation occurs. At higher temperatures $\gtrsim 77$ K, the scattering rates below 20 meV do not show a pronounced increase, and therefore the anomalous saturation regime does not occur. From these findings we may conclude that it is the energy-dependence of the acoustic phonon scattering rates at cryogenic temperatures which cause the saturation ``shoulder" in \textit{p}-Si between 10 and 200 \smalleunits~for $T\lesssim 40$ K, in agreement with a prior work \cite{VonB:1976}.

\subsection{Power spectral density}

\begin{figure}
    \centering{
        \includegraphics[width=3.4in, height=3.0in]{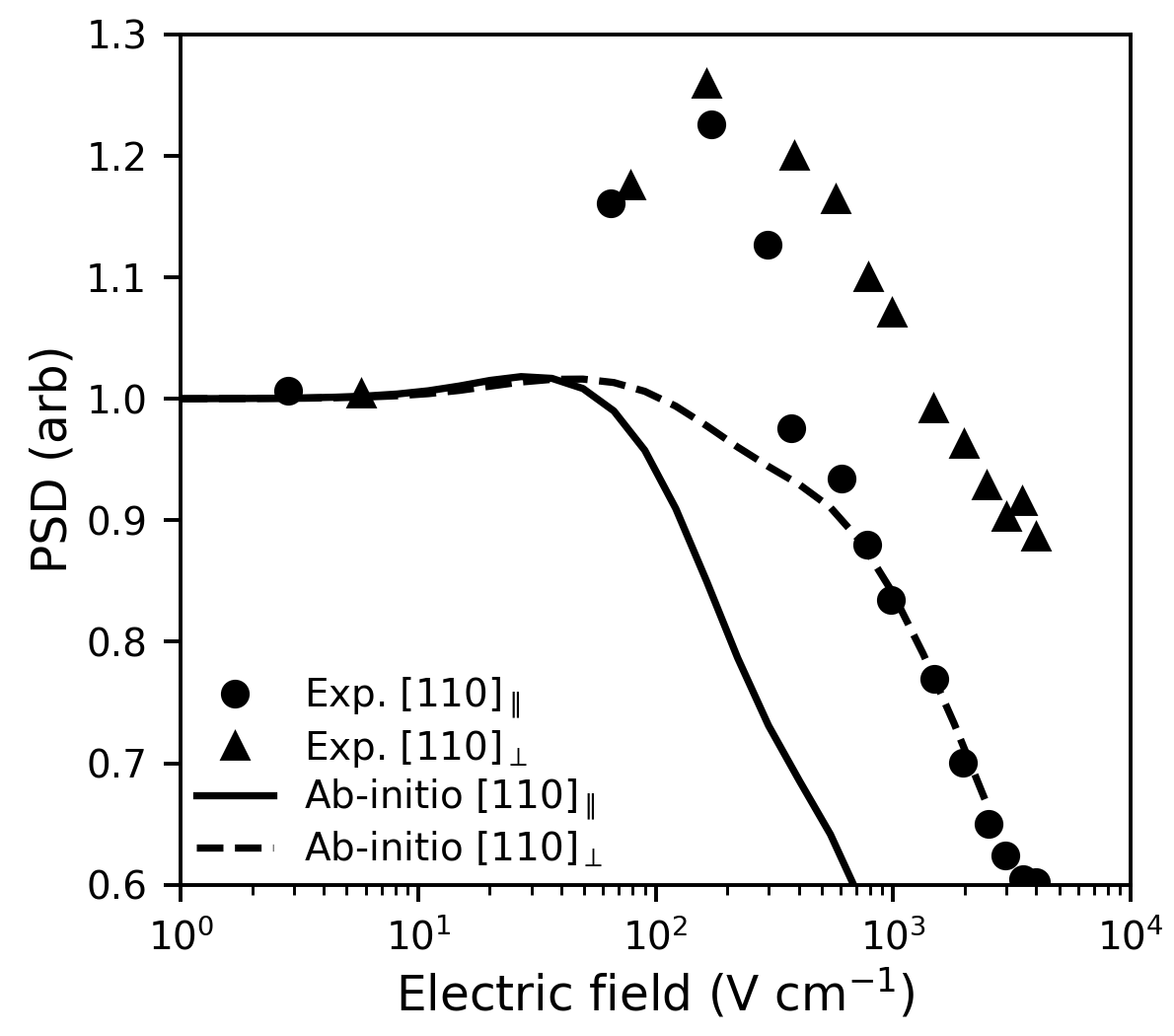}}
    \caption{PSD of current noise versus electric field applied in the [110] direction. Experimental values from Ref.~\cite{Bareikis:1981} (measurements taken parallel to [110] as upward triangles, measurements taken in the transverse direction [1$\overline{1}$0] as circles). Calculated data also shown (parallel ($\parallel$) as solid line, perpendicular ($\perp$) as dashed line).}
    \label{fig:PSD}
\end{figure}

We now turn to the non-monotonic trend of the microwave PSD versus electric field. This feature appears in the experimental 10 GHz PSD, which first increases and then decreases with an increasing electric field, leading to a peak at around 20 \smalleunits~\cite{Bareikis:1980,Bareikis:1981}. We have calculated the 10 GHz PSD for longitudinal and transverse measurement directions ([110] and [1$\overline{1}$0], respectively) versus [110] electric field strength, shown in \cref{fig:PSD}. The calculations do not agree quantitatively with experiment but show a qualitatively similar non-monotonic trend. The absence of quantitative agreement is not unexpected as discrepancies on the order of 50\% exist between Fig. 1b of \cite{Bareikis:1980} and the various other experimental reports of the diffusion coefficient, namely Fig. 2a of Ref.~\cite{Reggiani:1980} and Fig. 3 of Ref.~\cite{Gasquet:1986}.

Previous studies have attributed the trend to the competition of carrier heating (increasing noise) and decreasing relaxation time (decreasing noise) \cite{Bareikis:1981}, as well as the convective noise mechanism which is influenced by the field-dependent energy relaxation time \cite{Hartnagel:2001}. However, each explanation has inconsistencies. Regarding the first explanation, we note that the 10 GHz PSD measurements by Ref.~\cite{Bareikis:1980,Bareikis:1981} were used to calculate the diffusion coefficient, which is valid only in the low frequency limit ($\omega \tau \ll 1$, where $\tau$ is the average momentum relaxation time) \cite{Gantsevich:1979}. In this limit, an increase in transport properties with increasing electric field due to carrier heating would be expected in not just the PSD but also in DC quantities such as mobility. However, the hole mobility in Si is observed to decrease with increasing field. Therefore, this explanation is inconsistent with the observed trends of other transport properties. Regarding the second explanation, the convective noise mechanism only affects the longitudinal PSD, not the transverse PSD (see section 7.2 of Ref.~\cite{Hartnagel:2001}). However, because the PSD peak appears in both measurement directions, the convective mechanism cannot be responsible for the peak. Therefore, the underlying cause of the non-monotonic PSD feature remains unclear.\par

\begin{figure}
    \centering{
        \phantomsubcaption\label{fig:2D_plots_a}
        \phantomsubcaption\label{fig:2D_plots_b}
        \phantomsubcaption\label{fig:2D_plots_c}
        \includegraphics[width=7in, height=2.5in]{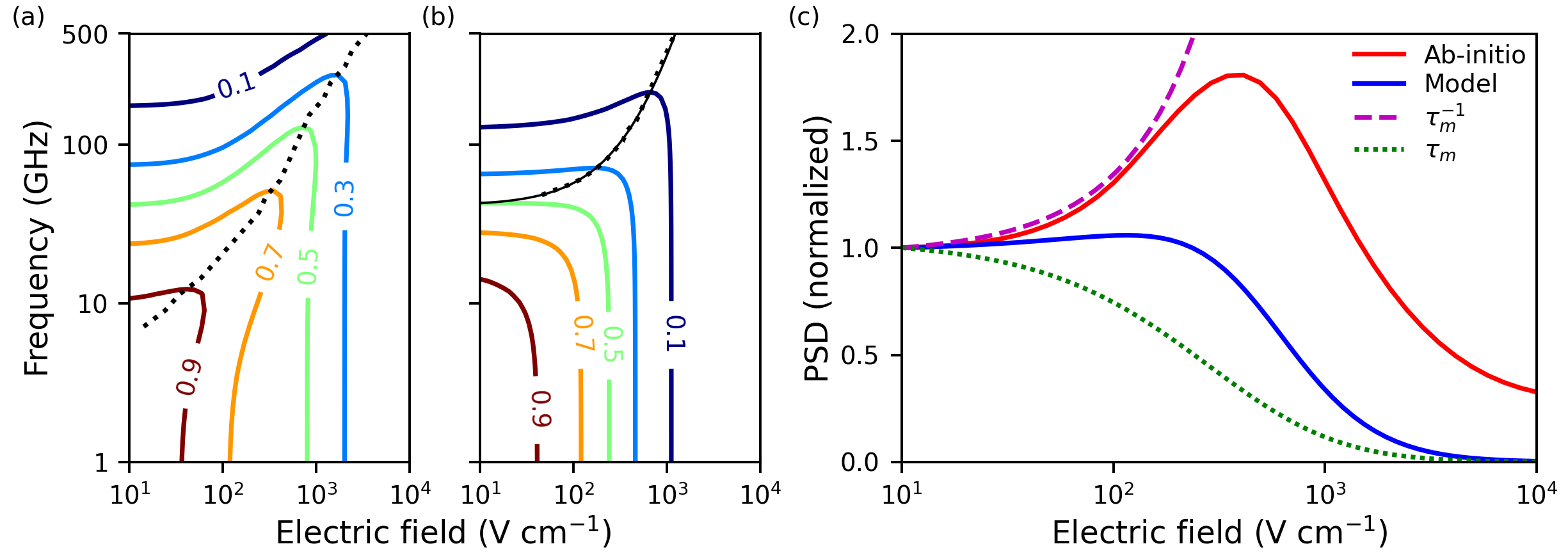}}
    \caption{(a) Contour map of normalized ab-initio PSD of current fluctuations versus frequency and electric field. Location of peak in the horizontal slices (PSD vs field curves) also shown (dotted black line) (b) Contour map of normalized model PSD (\cref{DrudePSD}) versus frequency and electric field. An analogous peak feature is observed as the ab-initio PSD. Location of peak in the horizontal slices shown (dotted black line), as well as $\omega\tau_m(E)=1$ (solid black line). (c) Ab-initio PSD vs. electric field in (a) (solid red line) and modelled PSD in (b) (solid blue line) at 60 GHz. The normalized modelled average momentum relaxation time is shown (dotted green line) as well as its inverse (dashed magenta line).}
    \label{fig:2D_plots}
\end{figure}

To identify the origin of the peak, we calculated the PSD versus both electric field and frequency. The result is shown in \cref{fig:2D_plots_a}. The longitudinal PSD calculation in \cref{fig:PSD} corresponds to a horizontal slice of this plot at 10 GHz. The location of the peak, given by the dotted line, is observed to depend on both electric field and frequency. Because the peak is observed in our calculations which include only the heavy hole band, interband scattering may be ruled out. Additionally, the feature is also observed when optical phonons are omitted, eliminating  carrier streaming noise as the origin (see Sec. 7.4 of Ref.~\cite{Hartnagel:2001}). Therefore, we conclude that the peak must arise from qualities of the hole and phonon dispersion and acoustic phonon scattering rates.\par

Knowing that the convective mechanism cannot be responsible for the PSD peak, and without any other mechanism by which energy relaxation may produce the feature, we take as a working hypothesis that the energy-dependence of the momentum relaxation time ($\tau_m$) leads to the trend. However, within the fully ab-initio framework, testing this hypothesis is difficult. Therefore, we introduce a simplified model relating the PSD to $\tau_m$. Assuming equipartition and neglecting anisotropy, the PSD takes the form \cite{Reggiani:1985}

\begin{equation}
    S_{j}(\omega,E)
    =
    \frac{S_{j}(0,E)} {1+\left(\omega\tau \right)^2}
\end{equation}

\noindent where $\tau$ represents a characteristic relaxation time of the system (for instance  momentum, energy, or intervalley relaxation), and $E$ represents the electric field strength. For our model, we assume a mono-energetic hole distribution with a phenomenological average momentum relaxation time depending only on electric field, $\tau_m(E)$. For simplicity, we can further approximate $S_{j}(0,E)\approx c\tau_m(E)$ as the low-frequency PSD is directly proportional to the average momentum relaxation time through the fluctuation-diffusion \cite{Gantsevich:1979} and Einstein \cite{Reggiani:1985_2} relationships. These assumptions yield

\begin{equation}
    S_{j}(\omega,E)
    \propto
    \frac{\tau_m(E)} {1+\left(\omega\tau_m(E) \right)^2}
    \label{DrudePSD}
\end{equation}

\noindent Now, the Einstein relationship holds only in the low-field limit, but it remains approximately valid at fields where the carrier distribution is sufficiently close to the equilibrium one. In the present case, the mean carrier energy at 20 \smalleunits, where non-monotonic behavior is apparent, is only 2.2\% larger than the equilibrium value, so the error arising from the use of the Einstein relationship should be on the order of this difference ($\lesssim 10\%$). Therefore, it is expected that \cref{DrudePSD} provides a qualitatively accurate description of PSD behavior at the electric fields of interest.

To plot \cref{DrudePSD}, an  expression for $\tau_m(E)$ is needed. From our ab-initio results, we related the momentum relaxation time to the mean carrier energy $\epsilon$ by fitting $\tau_m(\epsilon)=a\epsilon^{-b}$, while $\epsilon$ was related to the field strength by fitting $\epsilon=cE+\epsilon_0$, where $\epsilon_0$ is the mean carrier energy at equilibrium. From these fits we obtain $\tau_m(E)=3.86 \left(\epsilon(E)/\epsilon_0\right)^{-2.46}$, with $\tau_m$ in ps; and $\epsilon=1.9 \times  10^{-7} E + \epsilon_0$, where $\epsilon$ is in eV and $\epsilon_0=0.0133$ eV. Using this fit for $\tau_m(E)$, we compute PSD versus frequency and electric field strength using \cref{DrudePSD}. The result is given in \cref{fig:2D_plots_b}. We observe a qualitatively similar ``ridge" trend as seen in the ab-initio PSD. Horizontal slices, which yield PSD versus field strength, taken at high frequencies display a peak, the location of which follows the same trend as in \cref{fig:2D_plots_a}. Within this model, the location of the peak corresponds to $\omega\tau_m(E)=1$, providing evidence that the field-dependence of the average momentum relaxation time $\tau_m$ is responsible for the non-monotonic PSD trend.\par


We can understand this behavior by inspecting the relative magnitude of the numerator $\tau_m(E)$ and the denominator $1+\left(\omega\tau_m(E)\right)^2$, which represent carrier scattering and carrier inertia \cite{Champlin:1964}, respectively. At sufficiently high temperatures and low frequencies such that $\omega\tau_m(E) \ll 1$, the PSD is proportional to $\tau_m(E)$ and decreases with field due to $\tau_m(E)$ decreasing. However, when $\omega\tau_m(E) \gg 1$ is satisfied, which occurs at some combinations of sufficiently low temperatures, high frequencies, and low fields, then as $\tau_m(E)$ decreases the inertial term $1+\left(\omega\tau_m(E)\right)^2$ decreases faster than $\tau_m(E)$ itself. Thus, the PSD rises due to the resulting $\mathrm{PSD}(\omega,E)\sim1/\tau_m(E)$ relationship as carriers exhibit less inertia. Once $\omega\tau_m(E) \ll 1$ is satisfied again at high enough fields, which will occur regardless of temperature and frequency, the PSD again decreases with field. These different behaviors are displayed in \cref{fig:2D_plots_c}, along with horizontal slices of \cref{fig:2D_plots_a,fig:2D_plots_b} at 60 GHz. We have chosen this higher frequency for the plots to make the relevant features in the model more visible. We observe that at low fields, both the modelled and ab-initio PSD exhibit the expected $\tau_m^{-1}$ behavior, and the $\tau_m$ behavior at high fields. These different trends enable a peak in the PSD, driven solely by the field-dependence of the average momentum relaxation time.\par

We now discuss why the ab-initio PSD results display a peak even when $\omega\tau_m(E=0)\gtrsim 1$ is not satisfied at equilibrium, a condition strictly necessary for a peak to appear in the model. In a real system, the hole distribution is not characterized by a single energy, as was assumed in the model. Therefore, although the average momentum relaxation time $\tau_m(E)$ may satisfy $\omega\tau_m(E)\sim1$, the lifetime of some states $\tau_\qnumber = \Theta_{\qnumber, \qnumber}^{-1}$ may span a range of values of $\omega \tau_\qnumber$. Therefore, even at frequencies where $\omega\tau_m(E=0)<1$, some electronic states for which $\omega\tau_\qnumber\gg1$ may have non-negligible occupation. These states may then lead to a non-monotonic PSD trend even at lower frequencies than are required to satisfy $\omega\tau_m(E=0) \gtrsim 1$.



\section{Discussion}


We now examine our findings in the context of prior works. The role of the field-dependence of the average momentum relaxation time in producing non-monotonic trends of AC mobility with electric field has been previously reported \cite{Bonek:1972}. Various studies have incorporated field-dependence into the momentum relaxation time of AC mobility models and observed different trends depending on the frequency. At low frequencies such that $\omega\tau_m(E=0)\ll1$, the AC mobility is both theoretically predicted \cite{Gibson:1961,Paranjape:1961} and experimentally observed \cite{Glover:1973} to decrease monotonically with increasing electric field, as expected for systems with sublinear current-voltage characteristics. However, at frequencies such that $\omega\tau_m(E=0)\geq 1$, other studies predicted a non-monotonic dependence of AC mobility on electric field \cite{Mukhopadhyay:1969,Seeger:1973}. Experimentally, non-monotonic trends of AC mobility with electric field have been reported for \textit{n}-InSb \cite{Bonek:1970,Bonek:1972}. These experiments measured AC mobility at 77 K as a function of field from $0-165$ \smalleunits~ and from $0-83.7$ GHz, which includes the case $\omega\tau_m(E=0)\approx 1$. The measurements show non-monotonic behavior when $\omega\tau_m(E=0)>1$, with an initial rise in the AC mobility at low fields ($\lesssim 50$ \smalleunits), in contrast to the decreasing trend at lower frequencies. Furthermore, the data agree qualitatively with a hydrodynamic model \cite{Bonek:1970} which predicts non-monotonicity in the high-field AC mobility for frequencies above $\sim 50$ GHz. Since the AC mobility is proportional to the PSD at low electric fields through the Nyquist relationship, these prior findings are consistent with our conclusions. On the other hand, prior studies of the microwave PSD concluded the non-monotonic trend was due to a competition between carrier heating and decreasing average relaxation time with increasing field \cite{Bareikis:1981}, and convective noise \cite{Hartnagel:2001}. However, our study and prior AC mobility studies support the field-dependence of the average momentum relaxation time as the origin.

\section{Summary}

We have investigated the origin of anomalous high-field transport and noise characteristics in \textit{p}-Si at cryogenic temperatures using a modified high-field formalism for ab-initio charge transport. We find that the additional regime of drift velocity saturation that occurs for $T<40$ K is due to acoustic phonon emission, in agreement with a prior work. We also find that the peak in the power spectral density of current fluctuations which occurs at 77 K and 10 GHz is due to the energy-dependence of the momentum relaxation time, contrary to the conclusions of prior studies. This work highlights the capabilities of the ab-initio method for providing microscopic insight into high-field transport and noise phenomena in semiconductors.

\section*{Acknowledgements}

This work was supported by the National Science Foundation under Award No. 1911926.

\bibliography{bib}

\end{document}